\documentclass[10pt,prl,preprintnumbers,floatfix,aps,nofootinbib,showpacs,twocolumn,superscriptaddress,noshowpacs,preprintnumbers]{revtex4}

\usepackage{amsmath}
\usepackage{graphicx}
\usepackage[dvipsnames]{xcolor}
\usepackage{dcolumn}
\usepackage{bm}

\usepackage{url}
\usepackage{hyperref}
\usepackage{soul} 
\usepackage[normalem]{ulem} 
\usepackage{bm}
\usepackage{wasysym}
\usepackage{verbatim}
\usepackage{amsfonts}
\usepackage{ragged2e}

\definecolor{DARKBLUE}{HTML}{00008b}
\hypersetup{colorlinks=true,
urlcolor=DARKBLUE
,anchorcolor=DARKBLUE
,citecolor=DARKBLUE
,filecolor=DARKBLUE
,linkcolor=DARKBLUE
,menucolor=DARKBLUE
,linktocpage=true,
pdfproducer=medialab
}

\begin{document}
\preprint{IFT-UAM/CSIC-24-86}

\title{Enhancing Cosmological Model Selection with Interpretable Machine Learning
}
\author{Indira Ocampo}%
 \email{indira.ocampo@csic.es}
\author{George Alestas}
 \email{g.alestas@csic.es}
\author{Savvas Nesseris}%
 \email{savvas.nesseris@csic.es}

\affiliation{Instituto de Física Teórica UAM-CSIC, Universidad Autónoma de Madrid, Cantoblanco, 28049 Madrid, Spain.}

\author{Domenico Sapone}%
 \email{domenico.sapone@uchile.cl}

\affiliation{%
Departamento de Física, FCFM, Universidad de Chile, Santiago, Chile.}%

\date{\today}

\begin{abstract}

We propose a novel approach using neural networks (NNs) to differentiate between cosmological models, and implemented \texttt{LIME} as an interpretability approach to identify the key features influencing our model’s decisions.
We show the potential of NNs to enhance the extraction of meaningful information from cosmological large-scale structure data, 
based on current galaxy-clustering survey specifications, for the cosmological constant and cold dark matter ($\Lambda$CDM) model and the Hu-Sawicki $f(R)$ model. We find that the NN can successfully distinguish between $\Lambda$CDM and the $f(R)$ models, by predicting the correct model with approximately $97\%$ overall accuracy, thus demonstrating that NNs can maximize the potential of current and next generation surveys to probe for deviations from general relativity.
\end{abstract}

\maketitle

\label{sec:Intro}

\textbf{\emph{Introduction--}} The accelerated expansion of the Universe, successfully described by the cosmological constant $\Lambda$ and Cold Dark Matter ($\Lambda$CDM) model, presents significant enigmas due to recent tensions between low and high redshift observations, see \cite{Perivolaropoulos:2021jda} for a recent review. To account for the accelerated phase, alternative theories like covariant modifications of the Einstein-Hilbert action have been proposed \cite{Buchdahl:1970ldb, Starobinsky:1980te, Hu:2007nk, DeFelice:2010aj, Sotiriou:2008rp, Capozziello:2011et}. While these theories, namely the various covariant modifications to General Relativity (GR), provide potential solutions, they also face theoretical problems such as ghost-like behaviors or Ostrogradski instability \cite{Woodard:2006nt}, although the $f(R)$ models avoid these issues.

Lately, significant effort has been made to detect deviations from GR using data from early Universe physics like cosmic microwave background (CMB) photons from  \textit{Planck} \cite{Aghanim:2018eyx}, ACT \cite{ACT:2020gnv}, and the South Pole Telescope \cite{SPT:2014wbo, SPT:2018njh}, as well as from the distribution of baryonic matter at late times via BOSS \cite{BOSS:2016wmc, eBOSS:2020yzd, BOSS:2014hwf, BOSS:2013rlg}, DES \cite{DES:2021wwk, DES:2021esc}, DESI \cite{DESI:2024mwx, calderon2024desi, DESI:2023dwi, lodha2024desi}, and \textit{Euclid} \cite{EUCLID:2011zbd, Euclid:2021icp, Euclid:2024yrr}. These efforts have substantially reduced the uncertainty in cosmological parameters, however tensions remain in the standard $\Lambda$CDM model \cite{Abdalla:2022yfr}. Future surveys like the Simons Observatory \cite{SimonsObservatory:2018koc} and the Vera C. Rubin Observatory's LSST \cite{LSSTDarkEnergyScience:2018jkl} aim to further refine these measurements, forcing to have accurate theoretical modeling inducing extensive evaluations of the likelihood \cite{Cole:2021gwr}.

Given the complexities and limitations of cosmological model comparisons, which often rely on computationally intensive likelihood functions \cite{Kass:1995loi}, there is a pressing need to develop more sophisticated models to incorporate numerous nuisance parameters to capture astrophysical phenomena \cite{Euclid:2020tff, Euclid:2023tog, Bault:2024jet, Chen:2024tfp}. Here, we explore neural networks (NNs) as a Machine Learning (ML) tool, offering a new perspective in extracting meaningful information, especially in cases where there are degeneracies in the parameter space. The ability of ML to manage large data volumes and complex pattern recognition is demonstrated in its applications across astronomy \cite{Euclid:2024ofi,  Euclid:2024rko, Euclid:2024dru, Euclid:2024zgr, Euclid:2023uwd, Euclid:2023yvn, Euclid:2022ivm, Baqui:2020sfd, Bonjean:2019kuh, Mitra:2022nyc, YoungSupernovaExperiment:2022zqi, DES:2021rmp, DES:2022zrg} and high-energy physics \cite{Larkoski:2017jix, Guest:2018yhq, Kasieczka:2017nvn}. Moreover, an interesting approach in the case of building Bayesian emulators was performed in~\cite{Jeffrey:2023stk}, where the loss function was optimized to estimate the evidence. Other studies perform a Bayesian model comparison through neural classification with SN Ia data \cite{karchev2023simsims} and \cite{mancarella2022seeking, thummel2024beyond} to the matter power spectrum.

Here we also study the NN interpretability, in order to understand which are the relevant features (parameter values) that have a more significant impact in the decision making process. In particular, we perform this analysis using \texttt{LIME:} Local Interpretability Model Agnostic Explanations\footnote{\url{https://github.com/marcotcr/lime.git}} \cite{2016arXiv160204938T}, explaining individual predictions of the NN by creating perturbed samples around a specific instance, and evaluating how these local perturbations affect the result, and then showing which parts of the data were most important for that specific prediction. To demonstrate the advantage of our ML approach in distinguishing cosmological models, we focus on $\Lambda$CDM and the Hu-Sawicki $f(R)$ (HS) model \cite{Hu:2007pj, Hu:2007nk}.
We apply our method to simulated Stage-IV large-scale structure (LSS) survey data, specifically DESI-like surveys. This approach is timely given recent DESI data hinting at possible variations in the dark energy equation of state $w(z)$ \cite{DESI:2024mwx}, potentially indicating new physics or deviations from GR \cite{Tada:2024znt, Berghaus:2024kra, Ramadan:2024kmn}. However, special care should be taken when perturbations in the dark sector are considered \cite{Kunz:2006wc, Kunz:2006ca}.

\textbf{\emph{Setting the stage--}\label{sec:Theory_Data}} The HS model is described by the Lagrangian $R+f(R)$, 
where the function $f(R)$ for $|f_{R0}| \ll 1$ is approximately equal to \cite{Hu:2007nk}
\begin{equation}
f(R) = - 6\,\Omega_\mathrm{DE,0} \frac{H_0^2}{c^2} + |f_{R0}| \frac{\bar{R}_0^2}{R}+\ldots\,,\label{eq:fR}
\end{equation}
where $f_{R0}= {\rm d}f(R)/{\rm d }R|_{z=0}$ and then the background expansion history is well approximated by $\Lambda$CDM \cite{Basilakos:2013nfa}. Here we use $|f_{R0}|=  5\times 10^{-6}$, which is in agreement with observations; see for example, a summary of constraints in \cite{Euclid:2023tqw}. While \cite{Desmond:2020gzn} claims the constraint on $|f_{R0}|$ is tighter due certain observable signatures in galaxy morphology, it has been pointed out that such galactic-scale constraints are heavily dependent on the fine details of the screening mechanism assumed, see~\cite{Burrage:2023eol}.

Finally, in the context of the Friedmann--Lemaître--Robertson--Walker (FLRW) metric, the growth rate of the matter density perturbations $f\sigma_8$, can be defined as the logarithmic derivative of the matter density contrast.
Then, we can model the three-dimensional matter power spectrum, which measures the variance of the density contrast and is the Fourier transform of the 2-point correlation function \cite{Favole:2020ywr}. The observed galaxy power spectrum follows the modeling described in \cite{Euclid:2019clj}:
\begin{align}\label{eq:GC:pkobs}
P_\mathrm{obs}(k,\mu;z) =
& \frac{H(z)\,D_{\rm A,r}(z)}{H_{\rm r}(z)\,D_{\rm A}(z)}  
\left[b(z)\sigma_8(z)+f(z)\sigma_8(z)\mu^2\right]^2 \notag\\
& \times \frac{P_{\rm nl}(k,\mu;z)}{\sigma^2_8(z)} \, {\rm e}^{-k^2 \mu^2\sigma^2_r} + P_{\rm s}(z)\,,
\end{align}
where $\mu$ is the angle of the wave-vector $k$ with respect to the line of sight, and the $P_{\rm nl}$ is the matter power spectrum with non-linear corrections damping the BAO signal, see~\cite{Eisenstein:2006nj, Euclid:2023tqw}; the first term is the Alcock-Paczynski effect, which takes into account the change on volume via the Hubble parameter $H(z)$ and the angular diameter distance $D_{\rm A}(z)$ and it requires the assumption of a reference cosmology (subscript ``$r$'') to transform the observed redshifts into distances, while $P_{\rm s}(z)$ is the shot-noise term. The galaxy power spectrum is further modulated by an error in the redshift measurement $\sigma^2_r = c\,\delta z/H(z)$ with $\delta z = 0.0005\,(1 + z)$. The linear matter power spectrum for the models used in this work has been obtained using \texttt{MGCAMB}~\cite{Zhao:2008bn, Hojjati:2011ix, Zucca:2019xhg, Wang:2023tjj}. 

To simulate data, we assume a DESI-like survey covering $14,000,{\rm deg^2}$ targeting Bright Galaxies (BGS) at $z<0.4$, Luminous Red Galaxies (LRGs) at $0.4<z<1.1$, Emission Line Galaxies (ELGs) at $0.6 < z < 1.6$, and quasars at $0.9 < z < 2.1$ \cite{DESI:2016fyo, DESI:2023dwi}. We focus on BGS and ELG, distributing them across 16 redshift bins.

The covariance matrix is evaluated using the Fisher matrix approach for cosmological parameters such as $H(z)$, $D_{\rm A}(z)$, $f\sigma_8(z)$, and $b\sigma_8(z)$, keeping the shape parameters fixed to their reference values, i.e.  $\omega_\mathrm{m,0} = \Omega_\mathrm{m,0} = 0.2144$, $h = 0.67$, $\omega_\mathrm{b,0} = \Omega_{b,0} h^2 = 0.0335$, $n_s = 0.96$, and $\sigma_{8,0} = 0.85$, implying setting priors from CMB data \cite{Euclid:2021qvm}. The Fisher matrix has dimensions $16\,\text{bins} \times 5\,\text{parameters per bin} = 80$ parameters, marginalizing over nuisance parameters such as bias and shot noise. Regarding the $f\sigma_8(z)$, in any modified theory model or if dark energy perturbations are present, the growth rate depends on the scale $k$, see \cite{Nesseris:2015fqa}. Here, in order to constrain $f\sigma_8(z)$, we relaxed this assumption and took the value of the growth rate at $k=0.01\,\text{Mpc}^{-1}$, since the dependence across the $k$ range is less than $0.1\%$.

\begin{figure}
\centering 
\includegraphics[scale=0.16]{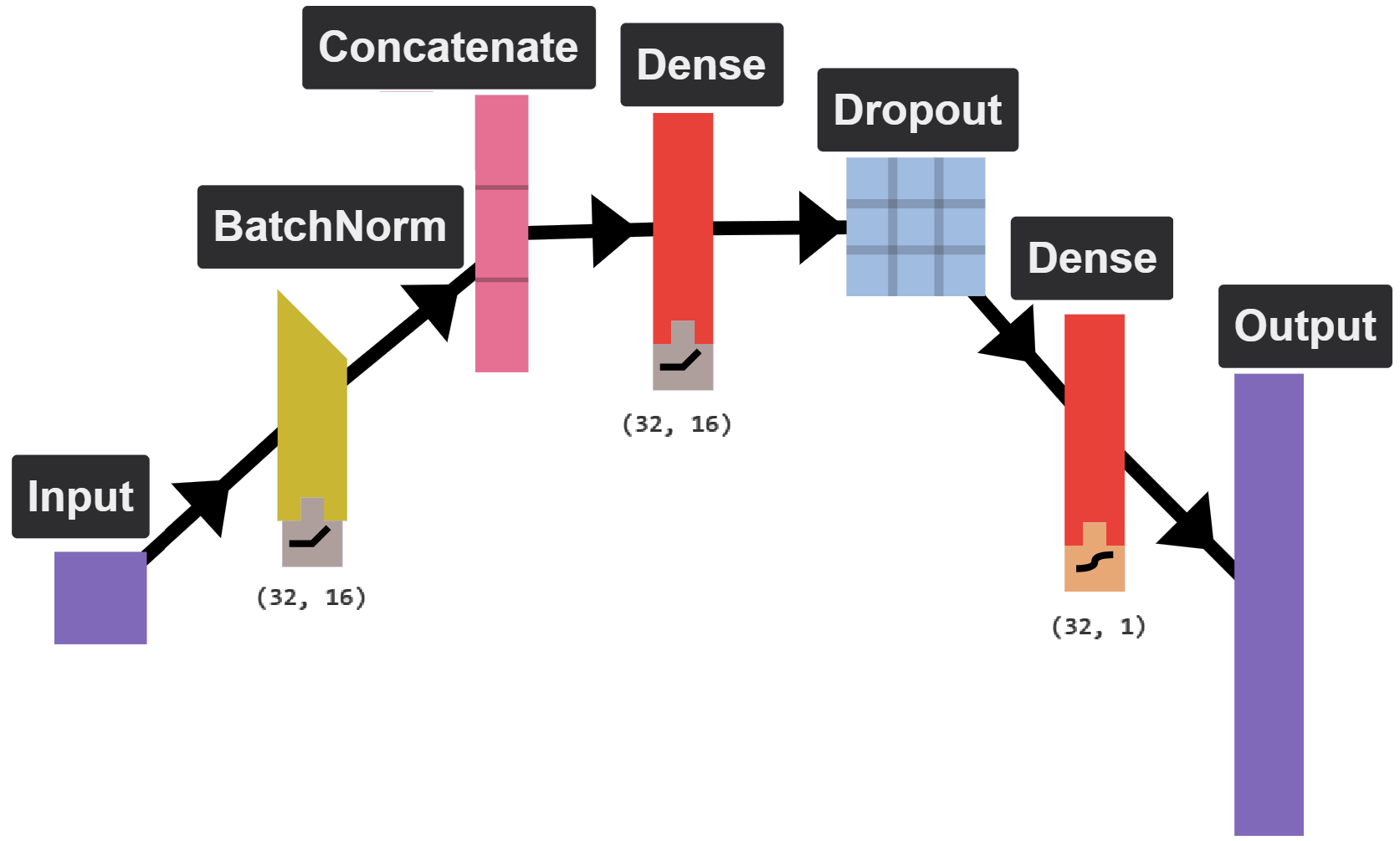}
\caption{\justifying Visualization of the architecture of our NN, where we implemented feature normalization with a \texttt{ReLU} activation function, followed by a concatenation layer. After that, the data passes through a fully connected layer also with a \texttt{ReLU} activation and a dropout layer, with a \texttt{dropout\_rate}$=0.2$, to finally go through another fully connected layer with a sigmoid activation function and assign a class.}
\label{fig:NN_architecture}
\end{figure}

\textbf{\emph{Simulated data--}}
In order to test deviations from $\Lambda$CDM with our methodology, we simulated DESI-like datasets for $f\sigma_8$ measurements reflecting both cosmological models: $\Lambda$CDM and HS. For this we varied the cosmological parameters in uniform rectangular grids, which are our priors, within the range: $\Omega_\mathrm{m,0} \in[0.2, 0.4]$ and $\sigma_{8,0} \in [0.7, 0.9]$ for $\Lambda$CDM, while for the HS model, we also vary $f_{R0}$ in the range $f_{R0}\in[10^{-6}, 5\times 10^{-6}]$. These values were chosen so that they are in agreement with current observational constraints. 

\textbf{\emph{NN architecture--}} 
We design a classifier which takes as input the simulated $f\sigma_8$ values for a DESI-like survey with 16 $z$-bins and returns as output the model: $\Lambda$CDM or Hu-Sawicki. For the simulated data, we introduced the uncertainties from a simulated DESI-like covariance matrix. The architecture implemented is shown in Fig.~\ref{fig:NN_architecture}, created with \texttt{ENNUI}.\footnote{\url{https://github.com/martinjm97/ENNUI.git}} We first implemented a feature normalization layer with a \texttt{ReLU} activation function \cite{agarap2018deep}, setting the \texttt{batch\_size} to 32 (for each of the 16 features). After normalization, we concatenated the features and passed the dataset through a fully connected layer, also with a \texttt{ReLU} activation function. Then, we applied a dropout layer, a regularization technique to prevent overfitting \cite{srivastava2014dropout}, with a dropout rate of 0.2. The final fully connected layer used a sigmoid activation.
Given the features $x$, the network approximates the probabilities for the classification task, HS: $p_{HS}= P(HS \mid x)/\left[P(HS \mid x) + P(\Lambda CDM \mid x)\right]$, or $\Lambda$CDM: $p_{\mathrm{\Lambda CDM}}=1-p_{\mathrm{HS}}$, these probabilities can be interpreted as evidence ratios, providing a measure of support for one model over the other. We also applied an early stopping callback \cite{ying2019overview} to prevent overfitting, setting the “patience” hyperparameter to 50 epochs. Our model achieved high accuracy and low loss in 1400 epochs. It was compiled with a nadam optimizer \cite{Dozat2016Feb}, a binary cross-entropy loss function and a learning rate of 0.001. The code will be available upon publication.\footnote{\url{https://github.com/IndiraOcampo/Growth_LSS_model_selection_Lime.git}}

\begin{figure}[t!]
\centering
\includegraphics[width = 0.495\textwidth]{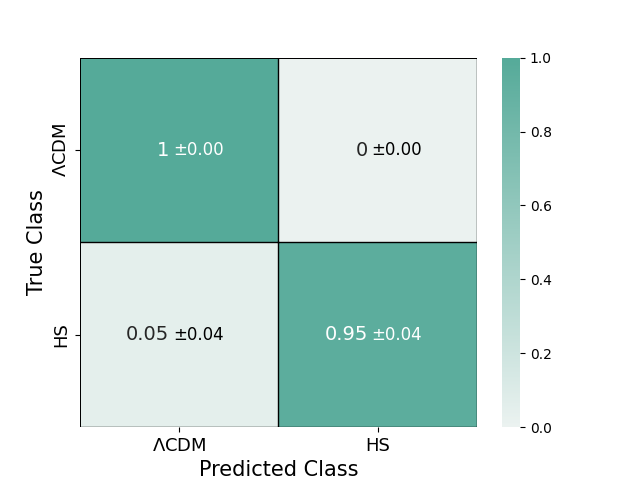}
\caption{\justifying The confusion matrix of the NN. As can be seen, the NN performs exceptionally well in discriminating the two models, predicting $\Lambda$CDM and the HS models with $100\%$ and $95\%$ accuracy, respectively. Note that the probabilities sum to unity horizontally.}
\label{fig:Conf_matrix}
\end{figure}

\textbf{\emph{Results.}\label{sec:Results}} 
The full dataset has 5000 $f\sigma_8$ samples (50$\%$ HS and 50$\%$ $\Lambda$CDM), that we split as $70\%$ for training + validation and 30$\%$ for testing. 
The NN has demonstrated well-converged loss and accuracy metrics showing minimal fluctuations. Finally, in Fig.~\ref{fig:Conf_matrix} we show the confusion matrix, where we notice that the NN performs very well, as it can correctly identify the $\Lambda$CDM model 100\% of the time and the HS model $\sim 95\%$ of the time, while offering false positives only in the rest $\sim 5\%$, these results in an overall accuracy rate of $97.5\%$ for a correct prediction.

\begin{figure}[t!]
\centering
\includegraphics[width = 0.5\textwidth]{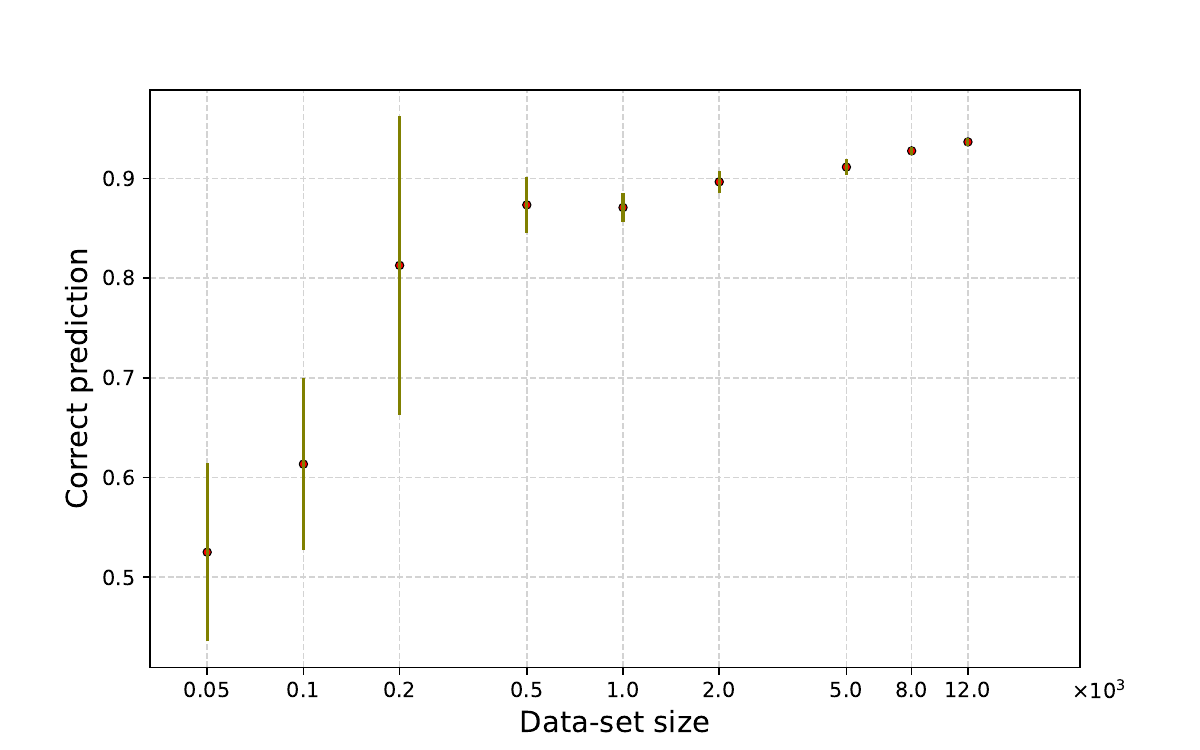}
\caption{\justifying Correct prediction performed by the neural network with respect to the number of dataset samples in semi-log scale, and the corresponding errors.}
\label{fig:dataset_size}
\end{figure}

\textbf{\emph{Robustness of the NNs--}\label{sec:Robust_NN}} 
We performed several tests to assess the robustness of our pipeline. First, we examined the effect of dataset size on NN performance, as shown in Fig.~\ref{fig:dataset_size}. The results indicate that correct predictions saturate after a few thousand realizations. Considering that the NN’s running time scales approximately linearly with the number of mock datasets, we chose a dataset size of 5000 realizations. This choice balances running time and NN accuracy.

We also investigated the NN’s performance using simulated datasets based on different covariance matrices. One matrix captured the inherent variations within the noise profile of the $\Lambda$CDM model, while the other represented the noise profile of the $f(R)$ model. We found that the NN displays an apparent accuracy of $100\%$ when using different covariance matrices, as the NN learns from the latter via the noise distribution of the $f\sigma_8$ data.

\begin{figure*}[!t]
\centering
\vspace{0.2cm}
\begin{minipage}{.48\textwidth}
\includegraphics[width=1.05\textwidth]{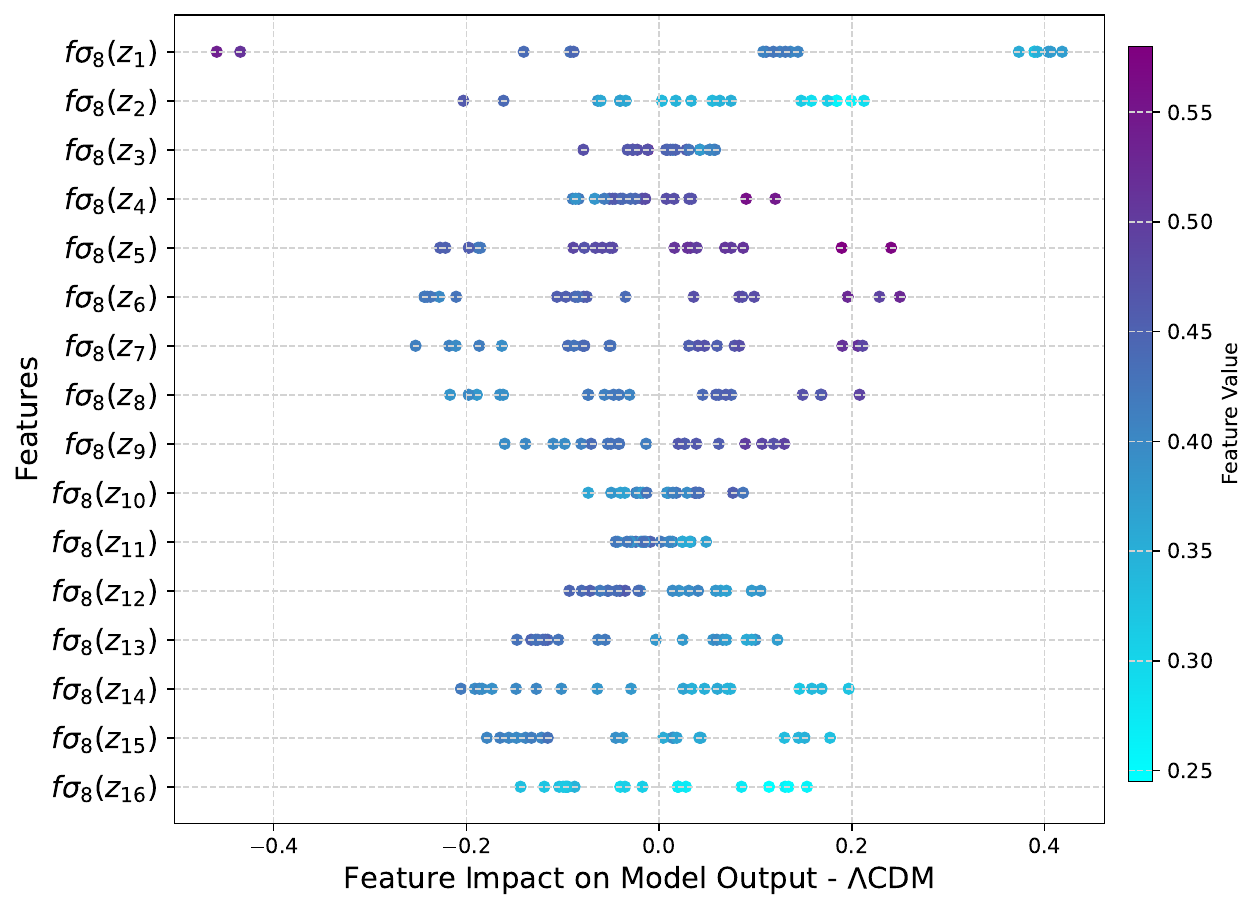}
\end{minipage}
\begin{minipage}{.48\textwidth}
\vspace*{-2cm}
\includegraphics[width=\textwidth]{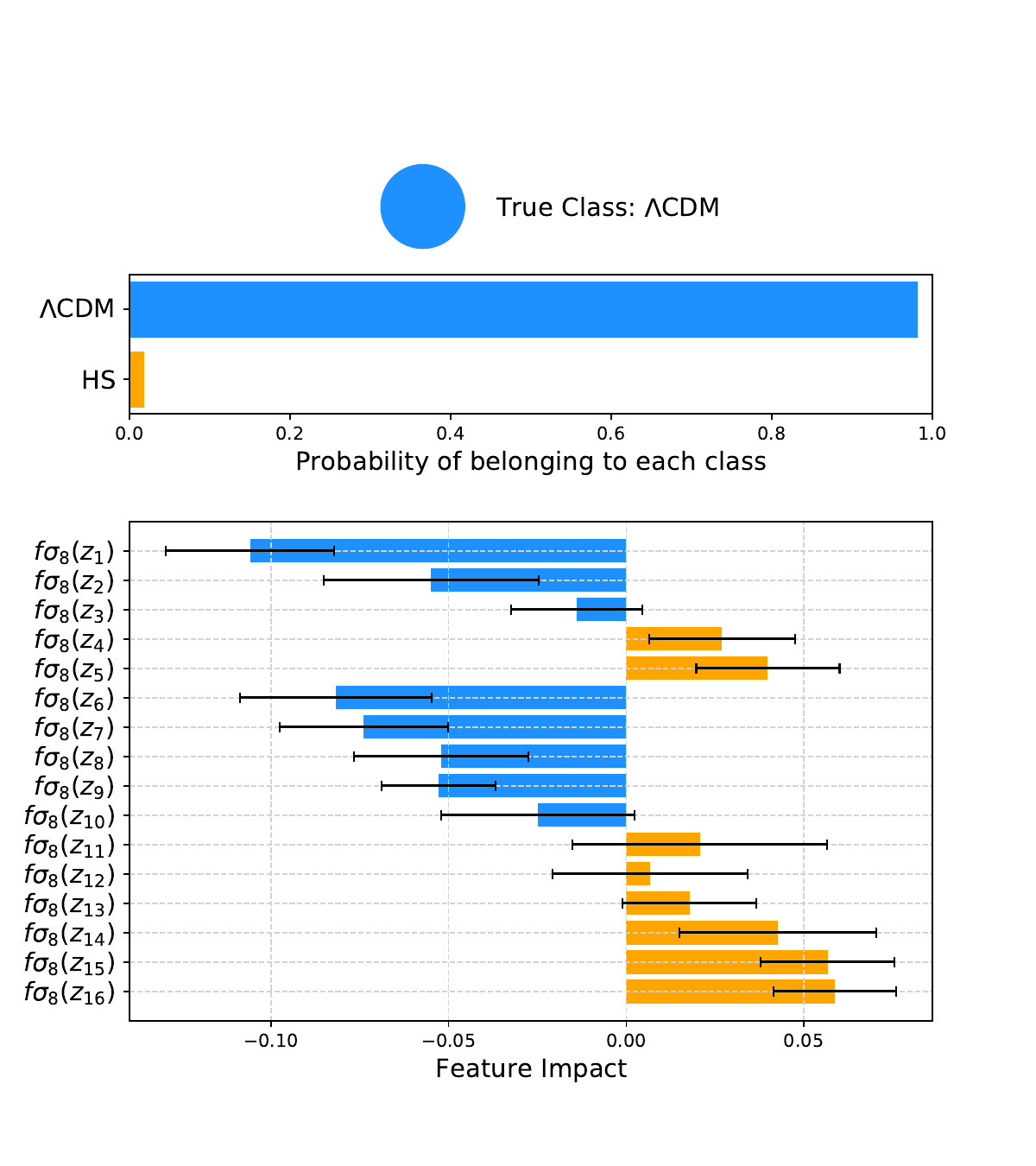}
\end{minipage}
\vspace*{-0.5cm}
\caption{\justifying Left: Impact of each feature ($f\sigma_8$ values) on the classification for 25 test examples correctly identified as $\Lambda$CDM, showing their influence on the final NN output: $\Lambda$CDM (negative $x$-axis) or HS (positive $x$-axis). Right: A single test example correctly classified, with (top) the overall probability of belonging to $\Lambda$CDM and (bottom) the individual feature probabilities influencing the final decision, with error bars.
}
\label{fig:LIME_LCDM}
\end{figure*}

\begin{figure*}[!t]
\centering
\begin{minipage}{.49\textwidth}
\includegraphics[width=1.05\textwidth]{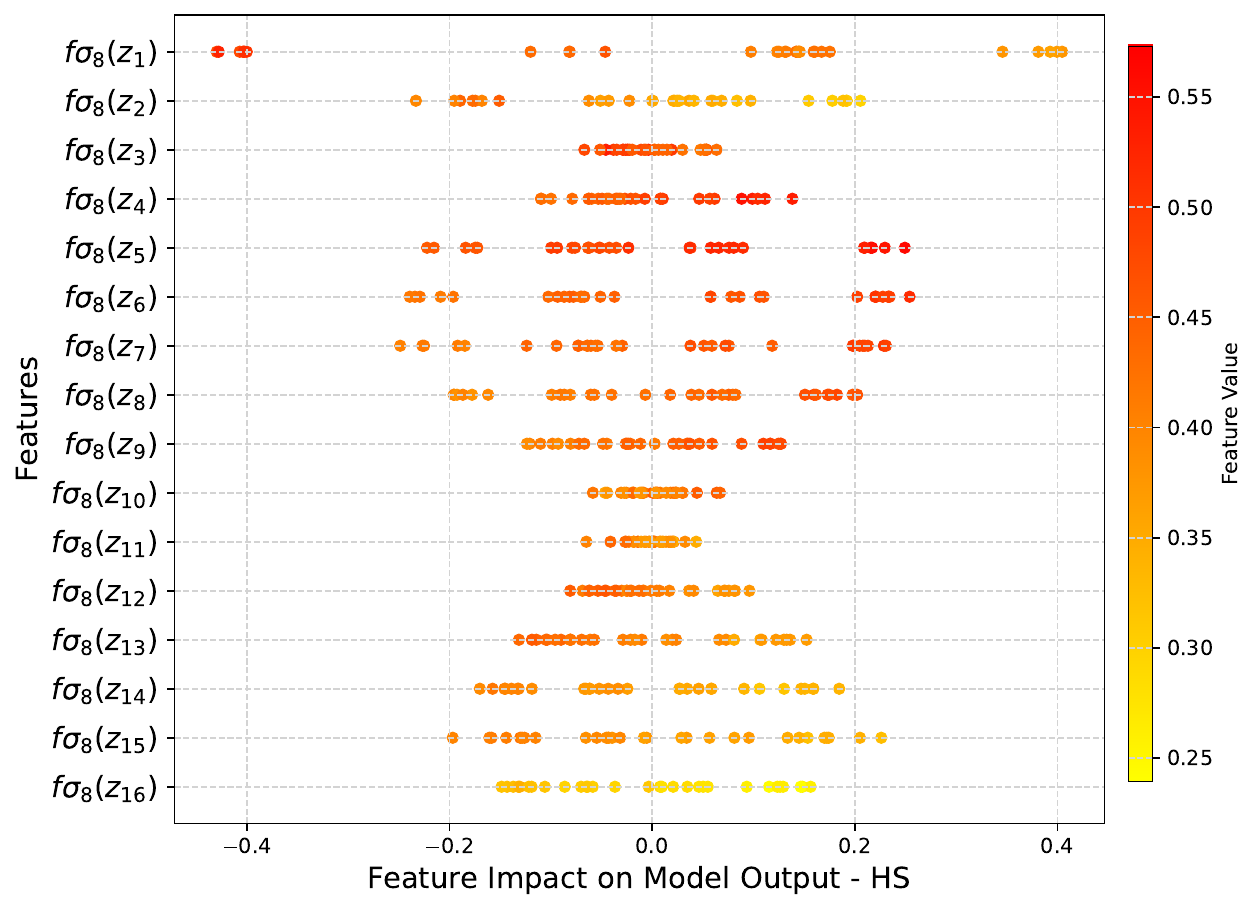}
\end{minipage}
\begin{minipage}{.49\textwidth}
\vspace*{-2cm}
\includegraphics[width=\textwidth]{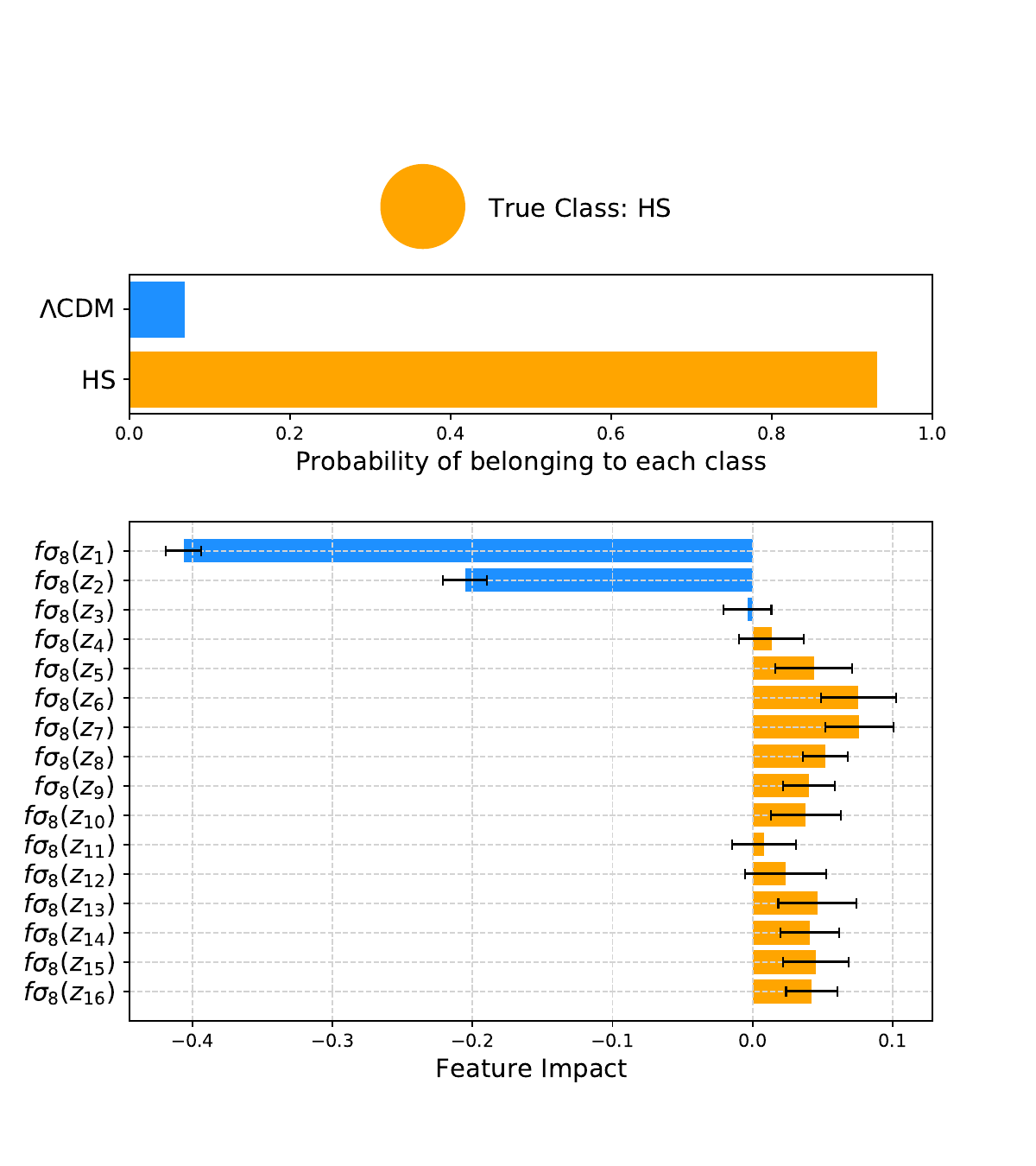}
\end{minipage}
\vspace*{-0.5cm}
\caption{\justifying As in Fig.~\ref{fig:LIME_LCDM}, but the correctly classified samples belong to the HS model.}
\label{fig:LIME_HS}
\end{figure*}

\textbf{\emph{NN interpretability, building local explanations--}} To identify which features drive the NN to discriminate between the two models, we implemented \textit{NN interpretability} \cite{NN_inter}. The first test involved training and testing with only the first and last eight $f\sigma_8$ values. The results showed that the first eight features predominantly influence performance. The scope of interpretable ML goes from local to global explanations, in this work we followed the first approach by implementing \texttt{LIME}. Local interpretability is simpler than global, aiming to understand the model’s decision-making process by generating nearby data points through random feature perturbations and analyzing their impact on predictions. 

We study the neighborhood around a selected sample in parameter space, using the trained model to classify data and assess the influence of each feature ($f\sigma_8$ input value) on the outcome \cite{lime, an2023specific}.
The right panels of Figs.~\ref{fig:LIME_LCDM} and ~\ref{fig:LIME_HS} display the impact of each $f\sigma_8$ value on the NN’s output for one data realization of the test set. Here \texttt{LIME} calculates individual feature impacts and determines the overall probability of a sample belonging to each class. The feature impact then is computed by \cite{ahern2019normlime, molnar2022}
\begin{equation}
\operatorname{Feature\;Impact}(x)=\arg \min _{g \in G} L\left(f, g, \pi_x\right)+\Omega(g),
\end{equation}

here, $g$ is the model that minimizes the loss function $L$, $\Omega(g)$ quantifies the model's complexity, which is kept low (e.g. by favoring fewer features), $G$ denotes the family of possible explanations, for example, all possible linear regression models. The proximity measure $\pi_x$ determines the size of the neighborhood around instance $x$ that is considered for the explanation. In practice, LIME focuses on optimizing the loss function, leaving the user to control the complexity, for instance by specifying the maximum number of features that the linear regression model can include.

For the correctly classified $\Lambda$CDM realization (Fig.~\ref{fig:LIME_LCDM} right), the NN assigns an overall $\Lambda$CDM probability of 98$\%$, although a few $f\sigma_8$ values favor HS (but with significant error bars). For the HS-classified sample (Fig.~\ref{fig:LIME_HS}), the NN assigns a 95$\%$ HS probability, with feature impacts generally supporting the correct class.

\begin{figure}
\centering 
\vspace{-0cm}
\includegraphics[scale=0.39]{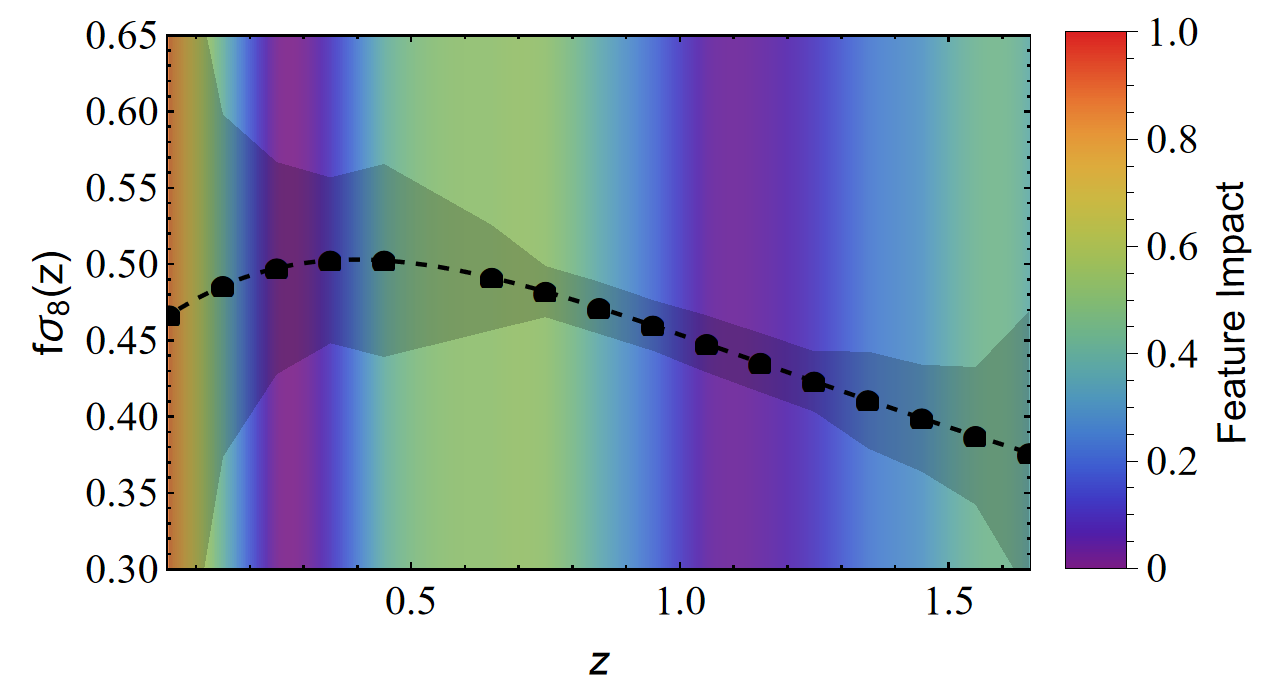}
\caption{\justifying One realization of $f\sigma_8$ data, as a function of the redshift $z$. The color shading corresponds to the feature importance of each redshift bin (blue tones correspond to lower feature importance and red to higher), according to the tests from \texttt{LIME}, and the gray colored region represents the errors.}
\label{fig:fs8_Feat_Imp}
\end{figure}

The left panels of Figs.~\ref{fig:LIME_LCDM} and ~\ref{fig:LIME_HS} show a generalized interpretability analysis of our NN, since they were computed with \texttt{LIME} for 50 data realizations of the test set (25 for each correctly classified model). We see the individual probability contributions of the features of each redshift bin, where $P(\mathrm{HS} \mid \mathbf{x}, \theta)$ are represented in the positive $x$-axis and $P(\Lambda \mathrm{CDM} \mid \mathbf{x}, \theta)$ in the negative one. Also, the $f\sigma_8$ values are displayed as a heat map. From this analysis, we extract the most important features that drive our NN to make decisions, and we display them in Fig.~\ref{fig:fs8_Feat_Imp}. Here we see the $f\sigma_8$ values across redshift $z$ with their corresponding feature importance emphasized using a color gradient shading scheme, transitioning from blue (low importance) to red (high importance). The most influential redshift bins are at low $(z<0.2)$, mid $(0.5<z<0.8)$, and high $(z>1.4)$ redshifts, where $\Lambda$CDM and HS models diverge the most. Conversely, intermediate redshifts $(0.2<z<0.5$ and $0.8<z<1.4)$ show very low importance, reflecting regions with lower growth rate differences.

\textbf{\emph{Conclusions--}} We have proposed a NN pipeline that successfully discriminates between the standard $\Lambda$CDM model, and the HS $f(R)$ model using LSS growth rate data. We also implemented an interpretability technique to understand which growth $f\sigma_8$ values are more relevant to the NN's output.
Overall, our pipeline achieves approximately $97\%$ accuracy in differentiating between models.

Our work focused exclusively on the galaxy-clustering $f\sigma_8$ data as a proof of concept to establish and demonstrate the method’s strengths. However, it can easily extend to use the multipoles of the redshift-space power spectrum and other related observables. We demonstrated the robustness of our approach by examining various aspects, such as the number of training samples and different covariance matrices for creating mock data. The novelty of our approach is the extraction of important information about the input features in the NN classification process.

To our knowledge, such a combination of observables used to directly test GR and a related interpretability NN pipeline has not been previously considered in the literature, while other tasks, like using NNs to speed up numerical calculations of evidence have been (see \cite{Chantada:2023vsy, Srinivasan:2024uax}). The potential to extend our pipeline is significant, especially with current LSS surveys. It can help discriminate models that are otherwise difficult to distinguish, thus opening new avenues to probe for deviations from GR, while it can also be useful to uncover the NNs learning process.

~\\
\textbf{\emph{Acknowledgements--}} We would like to thank G.~Ca\~nas, S.~Casas, L.~Goh, S. Farrens, E. Centofanti and V.~Pettorino for useful discussions. I. O. thanks ESTEC/ESA for the warm hospitality during the execution of this project, and for support from the ESA Archival Research Visitor Programme. I. O., G. A. and S. N. acknowledge support from the research project PID2021-123012NB-C43 and the Spanish Research Agency (Agencia Estatal de Investigaci\'on) through the Grant IFT Centro de Excelencia Severo Ochoa No CEX2020-001007-S, funded by MCIN/AEI/10.13039/501100011033. GA's research is also supported by the Spanish Attraccion de Talento Contract no. 2019-T1/TIC-13177 granted by the Comunidad de Madrid. I. O. is also supported by the fellowship LCF/BQ/DI22/11940033 from ``la Caixa” Foundation (ID 100010434). D. S. acknowledges financial support from the Fondecyt Regular project number 1200171.

\textbf{\emph{Data availability--}} The code is available at \cite{Growth_LSS_model_selection_Lime.git}, and the data at \cite{ocampo_2024_14515099}.

\bibliography{references}

\end{document}